# Mapping between the order of thermal denaturation and the shape of the critical line of mechanical unzipping in 1-dimensional DNA models

Sahin Buyukdagli and Marc Joyeux [(#)]

*Laboratoire de Spectrométrie Physique (CNRS UMR 5588),*

*Université Joseph Fourier - Grenoble 1,*

*BP 87, 38402 St Martin d'Hères, FRANCE*

**Abstract** : In this Letter, we investigate the link between thermal denaturation and mechanical unzipping for two models of DNA, namely the Dauxois-Peyrard-Bishop model and a variant thereof we proposed recently. We show that the critical line that separates zipped from unzipped DNA sequences in mechanical unzipping experiments is a power-law in the temperature-force plane. We also prove that for the investigated models the corresponding critical exponent is proportional to the critical exponent α, which characterizes the behaviour of the specific heat in the neighbourhood of the critical temperature for thermal denaturation.

[(#)] email : Marc.Joyeux@ujf-grenoble.fr

# 1 - INTRODUCTION

Several fundamental biological processes, like transcription and replication, involve the separation and recombination of the two strands that compose double-stranded (zipped) DNA molecules. Since transcription and replication are quite complex mechanisms, many early studies focused on thermal denaturation, that is the separation of the two strands upon heating [1-7]. In cells, unzipping is however not mediated by thermal activation but rather by proteins, which apply forces to separate and stretch complementary DNA strands. Compared to thermal denaturation studies, the recent mechanical unzipping experiments, where forces are applied to adjacent 5' and 3' strands of individual DNA molecules [8-16], therefore represent a great step towards the understanding of real biological processes. These experiments triggered in turn much theoretical effort (see [17-27] and references therein).

The purpose of the present paper is to point out that there actually exists a tight link between thermal denaturation and mechanical unzipping. Indeed, we will show that for the two 1-dimensional mesoscopic DNA models we investigated it is possible to establish a mapping between the order of the thermal denaturation phase transition and the shape of the critical line of mechanical unzipping in the temperature-force plane. More precisely, we will show that the critical line that separates zipped from unzipped sequences in mechanical unzipping experiments is a power-law, whose exponent is proportional to the critical exponent $\alpha$ that characterizes the behaviour of the specific heat in the neighbourhood of the critical temperature for thermal denaturation. This work therefore complements that of Singh and Singh, who showed that the critical force varies as the square root of the temperature gap to denaturation for the Dauxois-Peyrard-Bishop model with a large anharmonic stacking parameter $\rho$ [26].

The remainder of this paper is organized as follows. Sect. 2 provides a brief description of the two models used in this work, as well as the bases of the Transfer-Integral formalism. Calculation of the critical exponent α is next discussed in Sect. 3. In Sect. 4, we establish that the critical unzipping force evolves as a power of the temperature gap to denaturation, and that the corresponding critical exponent is related to α by a linear scaling law. Finally, we discuss the relevance of this study for real mechanical unzipping experiments and conclude in Sect. 5.

## 2 – MODELS AND TRANSFER-INTEGRAL CALCULATIONS

The potential energy of the two DNA models whose unzipping behaviour is studied in this paper is of the form

$$V = \sum_{n=1}^{N} (V_M(y_n) + W(y_n, y_{n-1})) \, , \tag{2.1}$$

where $y_n$ is the deviation from equilibrium of the distance between the bases of the $n^{\text{th}}$ pair. The one-particle Morse potential

$$V_M(y_n) = D(1 - e^{-a y_n})^2 \tag{2.2}$$

models the binding energy of the hydrogen bonds that connect paired bases. $W(y_n, y_{n-1})$ is a nearest-neighbor potential, which describes the stacking interaction between successive bases belonging to the same strand. The choice for $W(y_n, y_{n-1})$ is crucial, since the shape of the thermal denaturation transition, which is a collective effect, depends primarily on its form. The Dauxois-Peyrard-Bishop (DPB) model [28] assumes that the stacking interaction is of the form

$$W(y_n, y_{n-1}) = \frac{K}{2}(y_n - y_{n-1})^2 (1 + \rho e^{-\alpha(y_n + y_{n-1})}) \, . \tag{2.3}$$

The coupling constant of this interaction drops from $K(1+\rho)$ to $K$ as the paired bases separate, which decreases the rigidity of DNA sequences close to dissociation and results in a sharp first-order transition. Numerical values of the parameters used in this work are those of Ref. [26], that is $D=0.063$ eV, $a=4.2$ Å$^{-1}$, $K=0.025$ eV Å$^{-2}$, $\alpha=0.35$ Å$^{-1}$, except that we performed calculations with several values of $\rho$ ranging from 0 to 5.

The Joyeux-Buyukdagli (JB) model [29] instead assumes that

$$W(y_n, y_{n-1}) = \frac{\Delta H}{C}(1 - e^{-b(y_n - y_{n-1})^2}) + K_b(y_n - y_{n-1})^2 , \quad (2.4)$$

where the first term describes the finite stacking interaction and the second one the stiffness of the phosphate-sugar backbone. The sharpness of the melting transition predicted by the JB model is precisely due to the finite depth $\Delta H / C$ of the stacking interaction. In this work, we performed calculations with two sets of parameters, namely that of Refs. [29,30], that is $D=0.04$ eV, $a=4.45$ Å$^{-1}$, $\Delta H=0.44$ eV, $C=2$, $b=0.10$ Å$^{-2}$ and $K_b=10^{-5}$ eV Å$^{-2}$ (set JB-1), and that of Ref. [7], that is $D=0.048$ eV, $a=6.0$ Å$^{-1}$, $\Delta H=0.409$ eV, $b=0.80$ Å$^{-2}$ and $K_b=4\ 10^{-4}$ eV Å$^{-2}$ (set JB-2).

Transfer-Integral (TI) calculations are made possible by the fact that only nearest-neighbor interactions are considered in the potential energy of Eq. (2.1). One can therefore define a kernel

$$K(y_n, y_{n-1}) = \exp(-\frac{1}{k_B T}[\frac{1}{2}V_M(y_n) + \frac{1}{2}V_M(y_{n-1}) + W(y_n, y_{n-1})]) , \quad (2.5)$$

and write the partition function of a homogeneous DNA sequence of length $N$, whose first base pair is kept at a given distance $y_1 = y$ from equilibrium and whose end base pair is anchored (i.e. $y_N = 0$), in the form

$$Z(y) = \int K(y_2, y_1) K(y_3, y_2)...K(y_N, y_{N-1}) \times$$
$$\exp(-\frac{1}{2k_B T}[V_M(y_1) + V_M(y_N)])\delta(y_1 - y)\delta(y_N) dy_1 dy_2...dy_N \quad (2.6)$$

The basic trick of the TI method consists in expanding the kernel in an orthogonal basis

$$K(y_n, y_{n-1}) = \sum_k \lambda_k \Phi_k(y_n) \Phi_k(y_{n-1}) ,\qquad(2.7)$$

where the $\Phi_k$ and $\lambda_k$ are the eigenvectors and eigenvalues of the integral operator and satisfy

$$\int dx\, K(x, y) \Phi_k(x) = \lambda_k \Phi_k(y) .\qquad(2.8)$$

By substituting the kernel expansion of Eq. (2.7) in Eq. (2.6), one gets

$$Z(y) = \sum_k \lambda_k^{N-1} \Phi_k(0) c_k(y) ,\qquad(2.9)$$

where

$$c_k(y) = \Phi_k(y) \exp[-\frac{V_M(y)}{2k_B T}] .\qquad(2.10)$$

Similar considerations show that the partition function $Z_{\text{free}}$ of the sequence without external constraint (except for the last base pair that remains anchored) can be written in the form

$$Z_{\text{free}} = \sum_k \lambda_k^{N-1} \Phi_k(0) C_k ,\qquad(2.11)$$

where

$$C_k = \int c_k(y) dy .\qquad(2.12)$$

The work done in stretching the first base pair at distance $y$, $W(y)$, is the free energy difference between the stretched and the free chains, that is

$$W(y) = -k_B T \ln \frac{Z(y)}{Z_{\text{free}}} ,\qquad(2.13)$$

The derivative of $W(y)$ with respect to $y$ provides the average force $F(y)$, which is needed to keep the first base pair separated by $y$

$$F(y) = \frac{dW(y)}{dy} = -\frac{k_B T}{Z(y)} \frac{dZ(y)}{dy} .\qquad(2.14)$$

In the thermodynamic limit of infinitely long chains ($N \to \infty$), the sums in Eqs. (2.9) and (2.11) are dominated by the ground state of the TI operator with eigenvector $\Phi_0$ and eigenvalue $\lambda_0$. In this limit, the free energy per base pair, $f$, may therefore be obtained from

$$f = -k_B T \ln \lambda_0 \tag{2.15}$$

for both the stretched and free chains, and Eqs. (2.13) and (2.14) simplify to

$$W(y) = -k_B T \ln \frac{c_0(y)}{C_0}$$
$$F(y) = -\frac{k_B T}{c_0(y)} \frac{dc_0(y)}{dy} . \tag{2.16}$$

The critical force $F_c$ is defined as the average force, which is needed to keep the bases of the first pair at an infinite distance from one another, that is

$$F_c = F(y \to \infty) . \tag{2.17}$$

At that point, it must be emphasized that $F_c$ is usually substantially smaller than the force, which is actually needed to separate the two strands. This later force indeed corresponds to the maximum of $F(y)$ when $y$ is increased from zero to infinity. It turns out that there usually exists a very large force barrier at short $y$ separations (see for examples Figs. 5 and 6 of [26]). $F_c$ does not correspond to the value of $F(y)$ at the maximum of the barrier, but rather to the asymptotic value of $F(y)$ at large values of $y$.

At last, one might wish to write the free energy per base pair of the unstretched sequence, $f$, as the sum of a non-singular part, $f_{ns}$, and a singular part, $f_{sing}$,

$$f = f_{ns} + f_{sing} . \tag{2.18}$$

The non-singular part of the free energy, $f_{ns}$, should behave smoothly as temperature is raised up to the melting temperature $T_c$, while the singular part, $f_{sing}$, is expected to vary more sharply and remain constant once the two DNA strands are separated. Singularities at $T_c$ in the temperature evolution of the entropy and/or specific heat of the system should arise from $f_{sing}$ and not $f_{ns}$. The decomposition of Eq. (2.18) is not unique, but the requirement that $f_{sing}$ remains constant above $T_c$, that is when the two strands are widely separated, indicates that the most natural choice consists in considering that $f_{ns}$ is the free energy of non-interacting DNA single-strands, that is when the pairing potential $V_M(y_n)$ is omitted in Eq. (2.1). One consequently obtains

$$f_{ns} = -k_B T \ln z_{ss} ,  \qquad (2.19)$$

where

$$z_{ss} = \int \exp[-\frac{h_{ss}(u)}{k_B T}] du .  \qquad (2.20)$$

The function $h_{ss}(u)$ in Eq. (2.20) is equal to

$$h_{ss}(u) = D + \frac{K}{2} u^2  \qquad (2.21)$$

for the DPB model and to

$$h_{ss}(u) = D + K_b u^2 + \frac{\Delta H}{2}(1 - e^{-bu^2})  \qquad (2.22)$$

for the JB one. The integral in Eq. (2.20) can be evaluated analytically for the DPB model, leading to

$$f_{ns} = D - \frac{k_B T}{2} \ln(\frac{2\pi k_B T a^2}{K}) .  \qquad (2.23)$$

For the JB model, $f_{ns}$ can either be estimated numerically or, at the cost of a slight approximation, be computed from

$$f_{ns} = D - k_B T \ln[a(I_1(0) + I_2(0))] ,  \qquad (2.24)$$

where $I_1(0)$ and $I_2(0)$ are obtained by setting $F=0$ in the expressions for $I_1$ and $I_2$ in Eqs. (2.8) and (2.9) of Ref. [7]. Note that a factor $a$ was omitted in Eq. (2.7) of Ref. [7], which should actually read $g_u(T,F) = D - \ln[a(I_1+I_2)]/\beta$.

## 3 – DETERMINATION OF THE CRITICAL EXPONENT α

In this section, we calculate the critical exponent α of unconstrained DNA sequences described by the DPB and JB models. α is the exponent, which describes the behaviour of the specific heat per base pair, $c_V$, in the neighborhood of the critical temperature $T_c$, that is

$$c_V = -T\frac{\partial^2 f}{\partial T^2} \propto |t|^{-\alpha}, \qquad (3.1)$$

where $t$ is the reduced temperature $t = T/T_c - 1$. Strictly speaking, α should be computed from the singular part of $c_V$. Experimentalists, however, usually have no means to separate the singular from the non-singular part of the measured specific heat. Since the singular part is expected to vary much more sharply than the non-singular part in the neighbourhood of $T_c$, the usual approximation dating back to the pioneering work of Kadanoff et al [31] consists in estimating α from log-log fits of the evolution of $c_V$ close to $T_c$. This amounts to consider that "transients" arising from the non-singular part do not alter the estimation of α. This method is still used today with only minor modifications (see for example Refs. [32-34]). Determination of α with this method is illustrated in Fig. 2 of Ref. [30] and Fig. 10 of Ref. [7] for the JB model and sets of parameters JB-1 and JB-2, respectively. The obtained values of α (α=1.13 and α=1.33, respectively) are reported in the first column of Table 1. We performed TI calculations with the same grid of $y$ values as in Refs. [7,30] for the DPB model and four values of ρ ranging from 0.0 to 5.0. Results are illustrated in Fig. 1 and reported in the first column of Table 1. Note that in all these calculations, as well as in those discussed below, the critical temperature $T_c$ is obtained as the temperature for which the correlation length $\xi$, computed as in [7,29,30], is maximum.

We also determined α from the temperature evolution of the singular part of the free energy, $f_{sing}$, which is expected to vary according to

$$f_{sing} \propto |t|^{2-\alpha} , \qquad (3.2)$$

where $f_{sing}$ is obtained from Eqs. (2.15), (2.18) and (2.19). Fig. 2 shows the evolution of $\log_{10}(-f_{sing}/D)$ as a function of $\log_{10}(|t|)$ for the DPB model and four values of ρ ranging from 0.0 to 5.0, obtained with the same grid of *y* values as in Refs. [7,30]. The values of α deduced from these plots are reported in column (2) of Table 1. Also reported in the same column are the values of α obtained from similar calculations dealing with the JB model and sets of parameters JB-1 and JB-2.

Comparison of columns (1) and (2) of Table I indicates that both methods agree in predicting that, for the DPB model, α increases with ρ. This is due to the fact that, for the DPB model, the order of the melting phase transition decreases from 2 to 1 as ρ increases [26,35,36]. However, for both the DPB and the JB model, the values of α obtained from the temperature evolution of $c_V$ are systematically larger than those obtained from the evolution of $f_{sing}$. Moreover, the temperature evolution of $f_{sing}$ leads to "well-behaved" values of α, that is, to values that are systematically comprised between 0 and 1, while the values of α obtained from the evolution of $c_V$ are larger than 1 for the two sets of parameters of the JB model and for the DPB model with ρ=5.0. The reason for this is that the switching from second to first order phase transition causes the temperature evolution of the specific heat to depart from a power-law and approach a Dirac peak with infinite slope. This can be clearly seen in Fig. 3, which shows the temperature evolution of the entropy per base pair, $s = -\partial f/\partial T$, for increasing values of ρ. At last, one may note that additional calculations (not discussed here) show that the values of α obtained from the temperature evolution of $f_{sing}$ all satisfy the hyperscaling law $\alpha + \nu = 2$ (where ν is the critical exponent of the correlation length ξ), while values obtained from the temperature evolution of $c_V$ do not [7,30].

## 4 – THE LINEAR SCALING LAW RELATING σ TO α

In this section, we first show that for the models introduced in Sect. 2 the critical force $F_c(T)$ behaves according to a power-law in the neighbourhood of $T_c$, and then that the corresponding critical exponent σ is linked to α by a linear scaling law.

Fig. 4 shows the critical force $F_c(T)$ for the DPB model and four values of ρ ranging from 0.0 to 5.0, obtained from Eqs. (2.16) and (2.17) and the same grid of *y* values as in Refs. [7,30]. It is seen that the shape of the critical line $F_c = F_c(T)$ between zipped and unzipped sequences varies markedly with ρ : it indeed transforms from a straight line into a curve as the phase transition evolves from second order (small ρ) to first order (large ρ). Similar plots for the JB model and sets of parameters JB-1 and JB-2 can be found in Fig. 1 of Ref. [7]. At that point, it should be emphasized that the only models, which reproduce reasonably the experimentally observed thermal evolution of the critical force, are the JB model with set of parameters JB-2 (these parameters were precisely adjusted in order to reproduce the $F_c(T)$ curve [7]) and, to a somewhat lesser extent, the DPB model with $\rho = 5$. Nonetheless, the other models are used here for the purpose of comparison and to check the validity of the scaling law derived below.

Log-log plots, such as those shown in Fig. 5, further indicate that $F_c$ actually evolves as a power of *t* :

$$F_c \propto |t|^\sigma ,\qquad(4.1)$$

with a critical exponent σ that decreases from 1 (for ρ=0) to 1/2 (for large values of ρ). Note that the value $\sigma = 1/2$ was already reported by Singh and Singh for the DPB model with $\rho = 5$ [26]. The corresponding values of σ, as well as those obtained for the JB model and the JB-1 and JB-2 sets of parameters, are reported in column (3) of Table 1. Most interestingly, it is possible to relate the two critical exponents α and σ through a linear scaling law. This can be achieved by noticing that the free energy of the stretched unzipped sequence, $g_u$, writes, for the DPB model,

$$g_u = f_{ns} - \frac{F^2}{2K} , \qquad (4.2)$$

where $f_{ns}$ is given in Eq. (2.23). Since the critical force $F_c$ is such that $g_u = f$, comparison of Eqs. (2.18) and (4.2) shows that $F_c$ satisfies

$$f_{sing} = -\frac{F_c^2}{2K} . \qquad (4.3)$$

Similarly, for the JB model $g_u$ writes

$$g_u = D - k_B T \ln[a(I_1 + I_2)] , \qquad (4.4)$$

where the expressions for $I_1$ and $I_2$ can be found in Eqs. (2.8)-(2.9) of Ref. [7]. A little bit of algebra then shows that for sets of parameters JB-1 and JB-2 one has, to a good approximation

$$f_{sing} \approx -\frac{I_1(0)}{I_1(0) + I_2(0)} \frac{F_c^2}{4K_b} . \qquad (4.5)$$

By replacing Eqs. (3.2) and (4.1) in Eqs. (4.3) and (4.5), one immediately sees that for both models α and σ should consequently satisfy

$$2 - \alpha = 2\sigma . \qquad (4.6)$$

Comparison of columns (2) and (4) of Table 1 shows that this scaling law is, indeed, very well satisfied for both models.

At that point, it should be mentioned that other critical exponents for $F_c$ and/or other scaling laws involving σ have been derived for models that differ markedly from the ones studied here. For example, Bhattacharjee has shown that if stretched DNA is treated as two flexible interacting elastic strings tied together at one end, then the critical force $F_c$ evolves according to [18]

$$F_c \propto \left| v - v_c \right|^{1/|2-d|}, \qquad (4.7)$$

where v is the depth of the pairing potential (taken as a Dirac function) and $v_c$ the critical value of v at which mechanical unzipping occurs at a given temperature $T$. It is difficult to check the validity of Eq. (4.7) for the more complex DPB and JB models studied here, because for these models the order of the transition depends on geometrical factors not taken into account in [18], like for example the ratio of the widths of the pairing and stacking interactions [35].

On the other hand, it was shown that for the Poland-Scheraga model where self-avoiding interactions are accounted for (both within loops and between loops and the rest of the chain), the critical force scales like [37]

$$F_c \propto \left| t \right|^{\nu}. \qquad (4.8)$$

In Eq. (4.8), $\nu$ does not stand for the critical exponent of the correlation length ξ but rather the correlation length exponent of a self-avoiding random walk (the radius of gyration $R_G$ of a random walk of length $L$ scales as $L^{\nu}$). Numerically, $\nu$ is equal to 3/4 in dimension $d=2$ and to approximately 0.588 in dimension $d=3$. It therefore appears that the two models lead to different predictions.

At last, when stretched DNA is described as a self-avoiding walk on the three-dimensional Sierpinski gasket, one gets [38]

$$2 - \alpha = \frac{\sigma}{\nu_\theta} , \qquad (4.9)$$

where $\nu_\theta$ is the critical exponent of the end-to-end distance. Eq. (4.10) coincides with Eq. (4.6) when $\nu_\theta$ is assigned its mean-field value $\nu_\theta = 1/2$.

## 5 – CONCLUSION

We have shown that for both the DPB and JB models the critical force $F_c$ evolves as a power of $T_c - T$ and that the corresponding critical exponent, σ, is related to the critical exponent of the specific heat, α, through the linear relation $2 - \alpha = 2\sigma$. Quite interestingly, numerical calculations indicate that the values of α derived from σ according to this relation are in excellent agreement with the statistically relevant values of α determined from the evolution of the singular part of the free energy, $f_{\text{sing}}$. Mechanical unzipping experiments may therefore provide an accurate method to estimate the order of the thermal denaturation phase transition. One must however be conscious of three limitations.

First, real DNA is a heteropolymer and sequence inhomogeneity has marked effects on both thermal [39-46] and mechanical [8,9,13,15,16,23,47,48] unzipping. We showed in Ref. [45] that the essential effect of heterogeneity on thermal denaturation is to let different portions of the investigated sequences open at slightly different temperatures. We also pointed out that, besides this macroscopic effect, the local aperture of each portion is however very similar to that of a homogeneous sequence of the same length. Nonetheless, the precise effect of heterogeneity on the validity of the scaling relation of Eq. (4.6) still has to be investigated.

Moreover, mechanical unzipping experiments are not straightforward and the range of values of *t* spanned by today experiments [14,47] is not sufficiently broad to confirm the power-law evolution of the critical force.

At last, the theoretical result itself may be questioned. As discussed at the end of Sect. 4, different models for DNA unzipping may indeed lead to different critical behaviors and scaling laws. Moreover, even when focusing on Hamiltonian models expressed in terms of continuous dynamical variables, the obtained result depends essentially on the facts that (i) the free energy of stretched unzipped sequences, $g_u$, varies as the square of the applied force (see Eqs. (4.2) and (4.4)), and (ii) the singular part of the free energy of unstretched zipped sequences, $f_{\text{sing}}$, evolves as a power of $t$ in the neighbourhood of the critical temperature $T_c$. To the best of our knowledge, point (i) is relatively robust in the sense that, for all the models we are aware of, $g_u$ displays the same $F^2$ dependence. This includes of course the DPB (Eq. (4.2)) and JB (Eq. (4.4)) models, as well as the freely jointed chain model (Eq. (3) of Ref. [14]), but we checked that the same dependence also holds for the more involved helicoidal DNA model of Barbi et al [49]. In contrast, point (ii) seems to be more model-dependent. Indeed, while $f_{\text{sing}}$ evolves as a power of $t$ in the neighbourhood of $T_c$ for the DPB and JB models, this is no longer true for the helicoidal DNA model of Barbi et al [49]. Figs. 5 and 6 of Ref. [49] show that for this model $f_{\text{sing}}$ rather behaves as $\exp(-a/|t|)$ in the neighbourhood of $T_c$. This indicates that the helicoidal DNA model does not describe thermal denaturation as a phase transition but as a Kosterlitz-Thouless singularity. While there still exists, for this model, a mapping between the shape of the critical line of mechanical unzipping in the temperature-force plane and the sharpness of the thermal denaturation transition, this mapping no longer assumes the very simple form of Eq. (4.6). Further work, both experimental and theoretical, is needed to ascertain this point.

# TABLE CAPTION

**Table 1** : Values of the critical exponents α and σ obtained according to different methods described in the text and for different sets of parameters for the DPB and JB models. α is the critical exponent of the specific heat and σ that of the critical force $F_c$. Note that, according to Eq. (4.6), α and σ are related through $\alpha = 2(1-\sigma)$.

**FIGURE CAPTIONS**

**Figure 1** (color online) : Plot of $\log_{10}(c_V/k_B)$ as a function of $\log_{10}(|t|)$ for the DPB model and four values of ρ ranging from 0.0 to 5.0. The dash-dotted lines show the slopes from which the critical exponents α are estimated. These values of α are reported in column (1) of Table 1.

**Figure 2** (color online) : Plot of $\log_{10}(-f_{\text{sing}}/D)$ as a function of $\log_{10}(|t|)$ for the DPB model and four values of ρ ranging from 0.0 to 5.0. The dash-dotted lines show the slopes from which the critical exponents $2-\alpha$ are estimated. The corresponding values for α are reported in column (2) of Table 1.

**Figure 3** (color online) : Plot of the entropy per base pair $s$ (in units of the Boltzmann constant $k_B$) as a function of the reduced temperature $t$ for the DPB model and four values of ρ ranging from 0.0 to 5.0.

**Figure 4** (color online) : Plot of the critical force $F_c$ (expressed in pN) as a function of temperature $T$ for the DPB model and four values of ρ ranging from 0.0 to 5.0.

**Figure 5** (color online) : Plot of $\log_{10}(F_c)$ as a function of $\log_{10}(|t|)$ for the DPB model and four values of ρ ranging from 0.0 to 5.0. $F_c$ is expressed in pN. The dash-dotted lines show the slopes from which the critical exponents σ are estimated. These values of σ are reported in column (3) of Table 1.

TABLE 1

|  | α (1) | α (2) | σ (3) | 2(1-σ) (4) |
|---|---|---|---|---|
| DPB, ρ=0.0 | 0.00 | 0.02 | 0.99 | 0.02 |
| DPB, ρ=0.5 | 0.58 | 0.35 | 0.78 | 0.44 |
| DPB, ρ=1.0 | 0.92 | 0.66 | 0.67 | 0.66 |
| DPB, ρ=5.0 | 1.53 | 0.99 | 0.50 | 1.00 |
| JB-1 | 1.13 | 0.82 | 0.59 | 0.82 |
| JB-2 | 1.33 | 0.57 | 0.72 | 0.56 |

(1) obtained from the plot of $\log_{10}(c_V/k_B)$ as a function of $\log_{10}(|t|)$ (see Fig. 1).

(2) obtained from the plot of $\log_{10}(-f_{sing}/D)$ as a function of $\log_{10}(|t|)$ (see Fig. 2).

(3) obtained from the plot of $\log_{10}(F_c)$ as a function of $\log_{10}(|t|)$ (see Fig. 5).

(4) compare with the values of α in column (2).

FIGURE 1

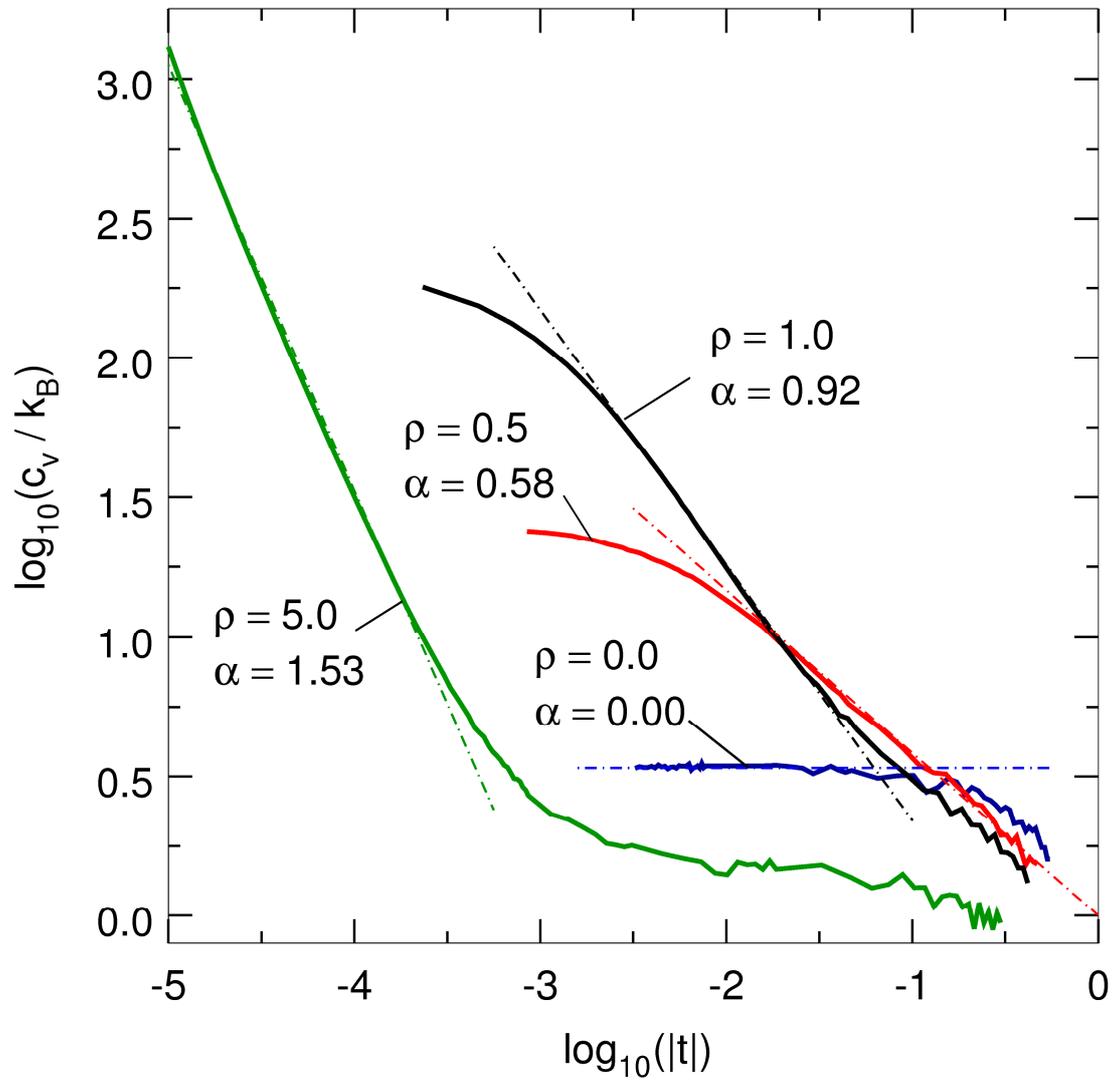

FIGURE 2

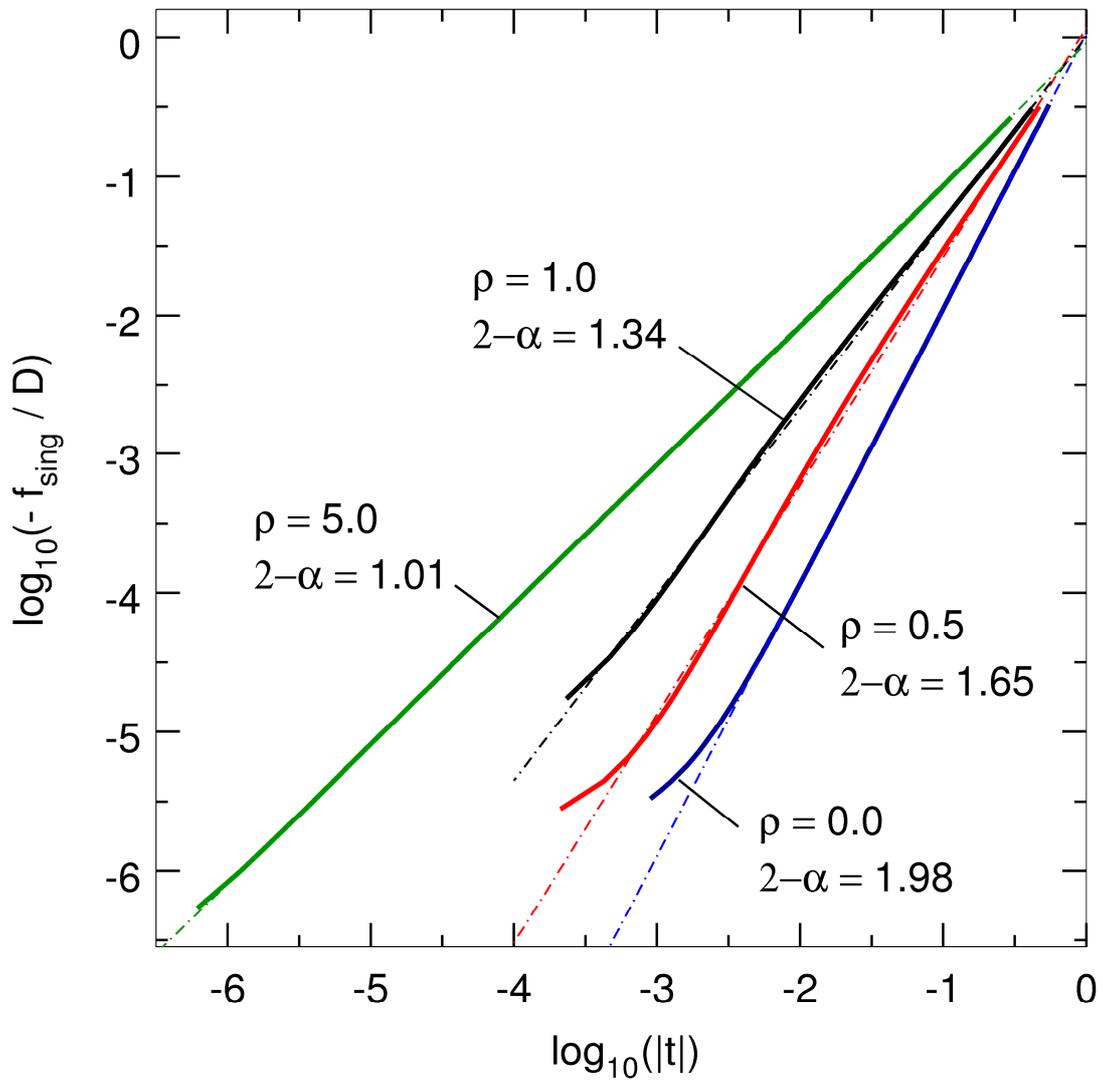

FIGURE 3

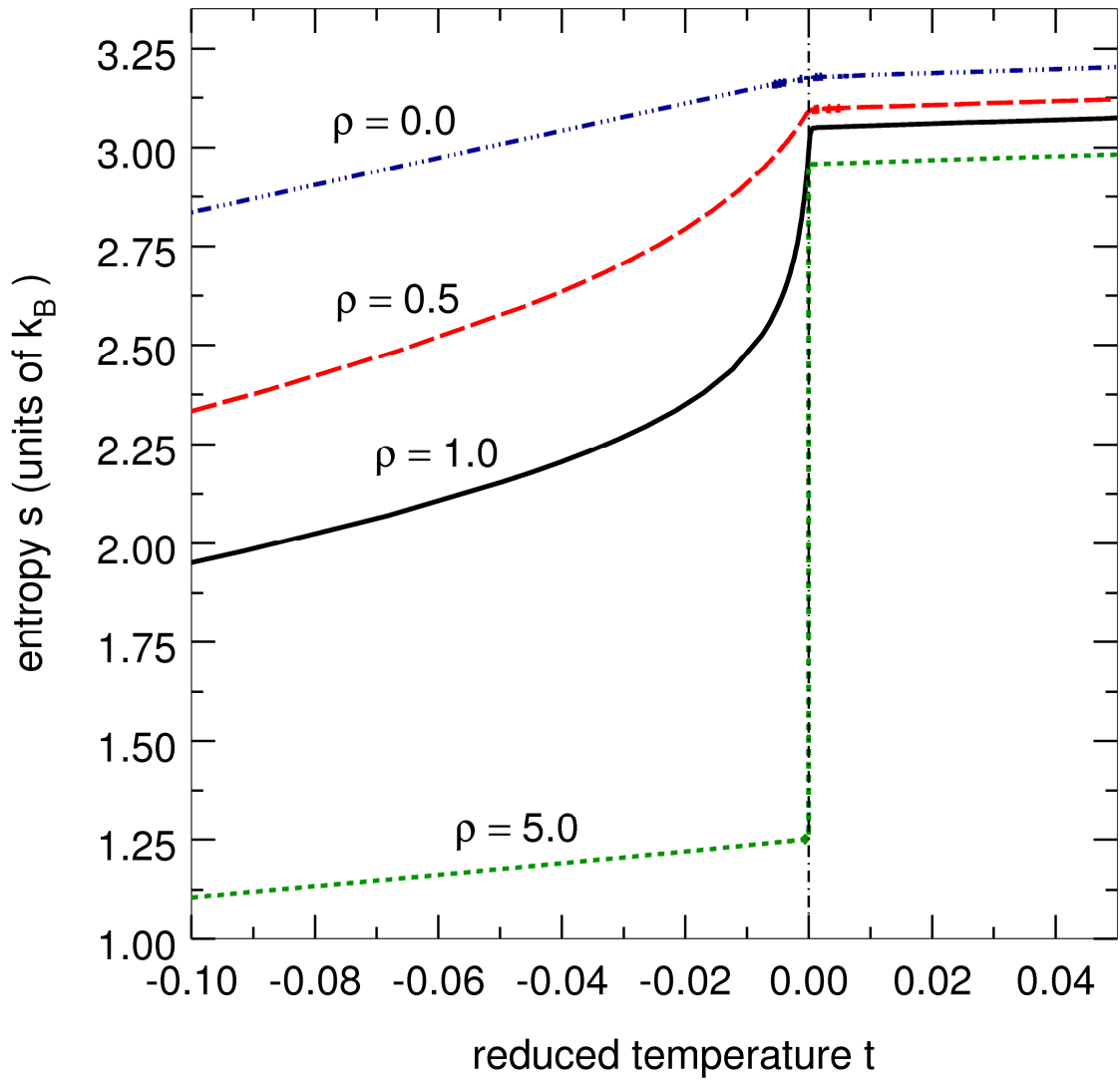

FIGURE 4

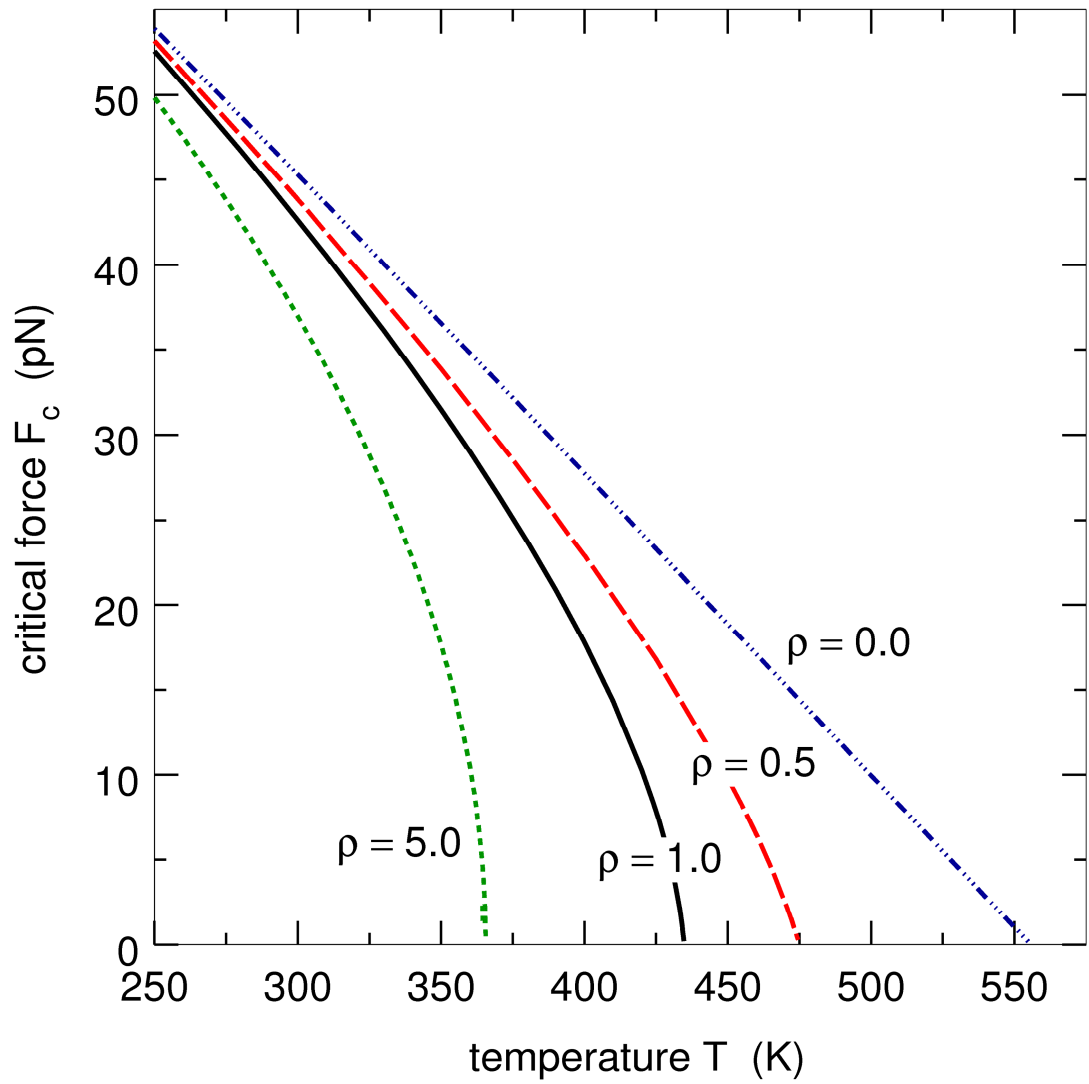

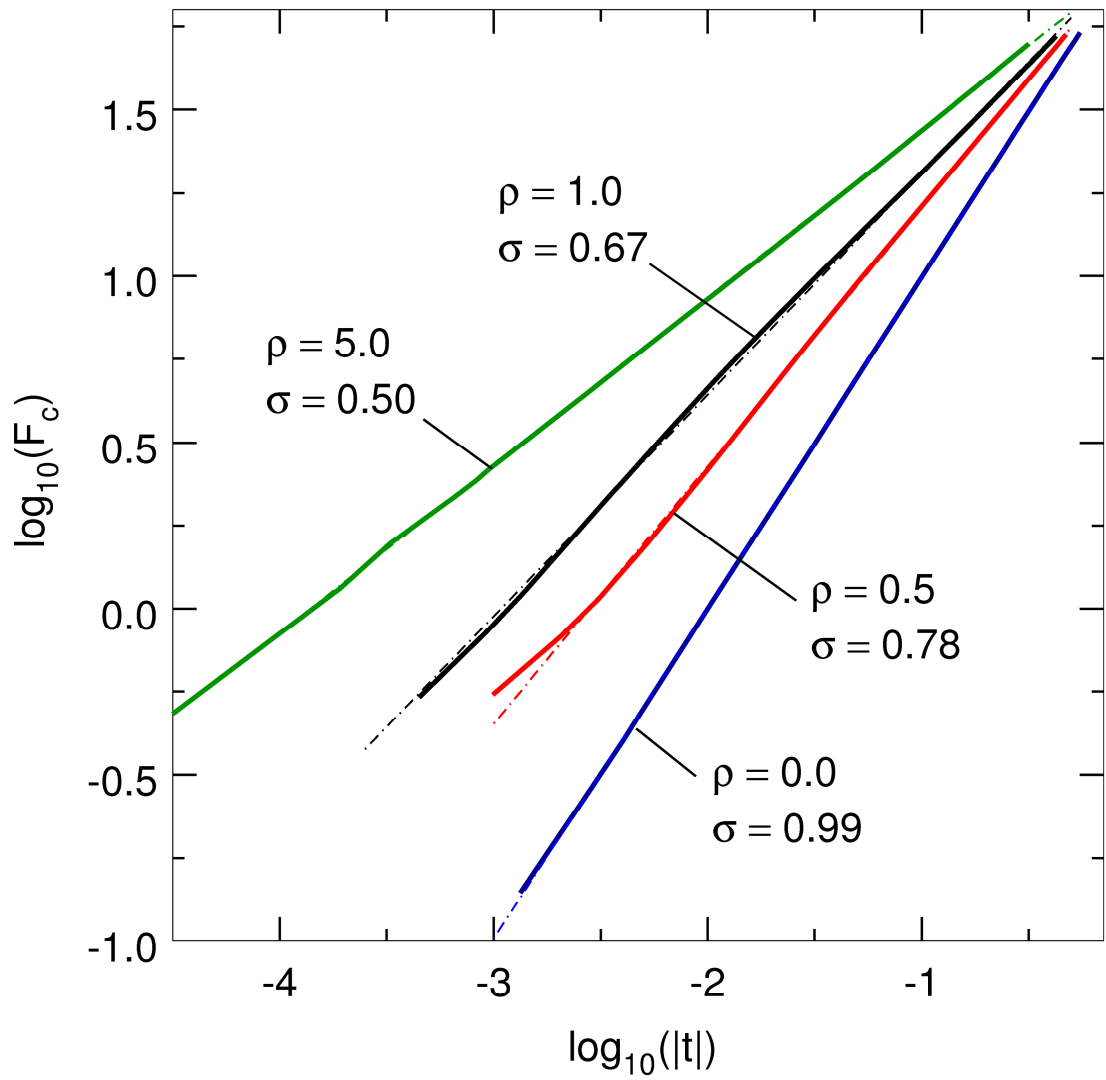